\documentclass[aps,prl,twocolumn,floatfix,superscriptaddress]{revtex4}

\usepackage{amsmath}
\usepackage{amssymb}
\usepackage{times}
\usepackage{braket}
\usepackage[pdftex]{graphicx}
\usepackage{color}

\renewcommand{\vec}[1]{\boldsymbol{#1}}
\newcommand{\change}[1]{\textcolor{black}{#1}}
\newcommand{\changeb}[1]{\textcolor{black}{#1}}

\newcommand{\sublattice}[1]{\ensuremath{\mathsf{#1}}}

\begin{document}

\title{Antiferromagnetic skyrmion crystals: generation, topological Hall and topological spin Hall effect}

\author{B{\"o}rge G{\"o}bel}
\email[]{bgoebel@mpi-halle.mpg.de}
\affiliation{Max-Planck-Institut f\"ur Mikrostrukturphysik, D-06120 Halle (Saale), Germany}

\author{Alexander Mook}
\affiliation{Max-Planck-Institut f\"ur Mikrostrukturphysik, D-06120 Halle (Saale), Germany}

\author{J\"urgen Henk}
\affiliation{Institut f\"ur Physik, Martin-Luther-Universit\"at Halle-Wittenberg, D-06099 Halle (Saale), Germany}

\author{Ingrid Mertig}
\affiliation{Max-Planck-Institut f\"ur Mikrostrukturphysik, D-06120 Halle (Saale), Germany}
\affiliation{Institut f\"ur Physik, Martin-Luther-Universit\"at Halle-Wittenberg, D-06099 Halle (Saale), Germany}

\date{\today}

\begin{abstract}
Skyrmions are topologically nontrivial, magnetic quasi-particles, that are characterized by a topological charge. A regular array of skyrmions---a skyrmion crystal (SkX)---features the topological Hall effect (THE) of electrons, that, in turn, gives rise to the Hall effect of the skyrmions themselves. It is commonly believed that \emph{antiferromagnetic} skyrmion crystals (AFM-SkXs) lack both effects. In this \changeb{Rapid Communication}, we present \change{a generally applicable method to create stable AFM-SkXs by growing a two sublattice SkX onto a collinear antiferromagnet. As an example we  show} that both types of skyrmion crystals---conventional and antiferromagnetic---exist in honeycomb lattices. While AFM-SkXs with equivalent lattice sites do not show a THE, they exhibit a topological spin Hall effect. On top of this, AFM-SkXs on inequivalent sublattices exhibit a nonzero THE, which may be utilized in spintronics devices. Our theoretical findings call for experimental \change{realization}.
\end{abstract}

\maketitle

\paragraph*{Introduction.}
\change{Skyrmions~\cite{skyrme1962unified,bogdanov1989thermodynamically, bogdanov1994thermodynamically,rossler2006spontaneous,muhlbauer2009skyrmion} are small magnetic quasiparticles, which are} usually caused by the Dzyaloshinskii-Moriya interaction~\cite{dzyaloshinsky1958thermodynamic,moriya1960anisotropic}, \change{but they} have been produced by other mechanisms~\cite{nagaosa2013topological}, \change{like frustrated exchange interactions~\cite{okubo2012multiple}}, as well. \change{While single skyrmions are envisioned to be used as ``bits'' in data storage devices~\cite{fert2013skyrmions,wiesendanger2016nanoscale,romming2013writing,hsu2016electric, zhang2015magnetic,zhang2015magnetic2,jiang2015blowing,boulle2016room,seki2012observation,woo2016observation}, which provide durability of data due to topological protection~\cite{nagaosa2013topological}, skyrmion \emph{crystals} (SkXs)---regular arrays of skyrmions---are best known for exhibiting the topological Hall effect (THE) of electrons~\cite{neubauer2009topological,schulz2012emergent,kanazawa2011large, lee2009unusual,li2013robust,bruno2004topological,gobel2017THEskyrmion, gobel2017QHE,ndiaye2017topological}, that, in turn, gives rise to the skyrmion Hall effect (SkHE; also present in isolated skyrmions)~\cite{nagaosa2013topological,zang2011dynamics,jiang2017direct,litzius2017skyrmion}.} 

\change{From the perspective of applications in data storage devices, the SkHE is undesirable. Thus, the concept of antiferromagnetic (AFM) skyrmions has been developed~\cite{barker2016static,zhang2016antiferromagnetic,PhysRevB.95.054421,jin2016dynamics}: skyrmions on two sublattices in which the spins on one sublattice are (mutually) reversed. As a result, both THE and SkHE vanish~\cite{barker2016static}. Because no periodic antiferromagnetic skyrmion crystal (AFM-SkX) is known yet, surrogate systems consisting of two skyrmion layers with opposite winding have been investigated~\cite{zhang2016magnetic,buhl2017topological}.}  

\begin{figure}
  \centering
  \includegraphics[width=.9\columnwidth]{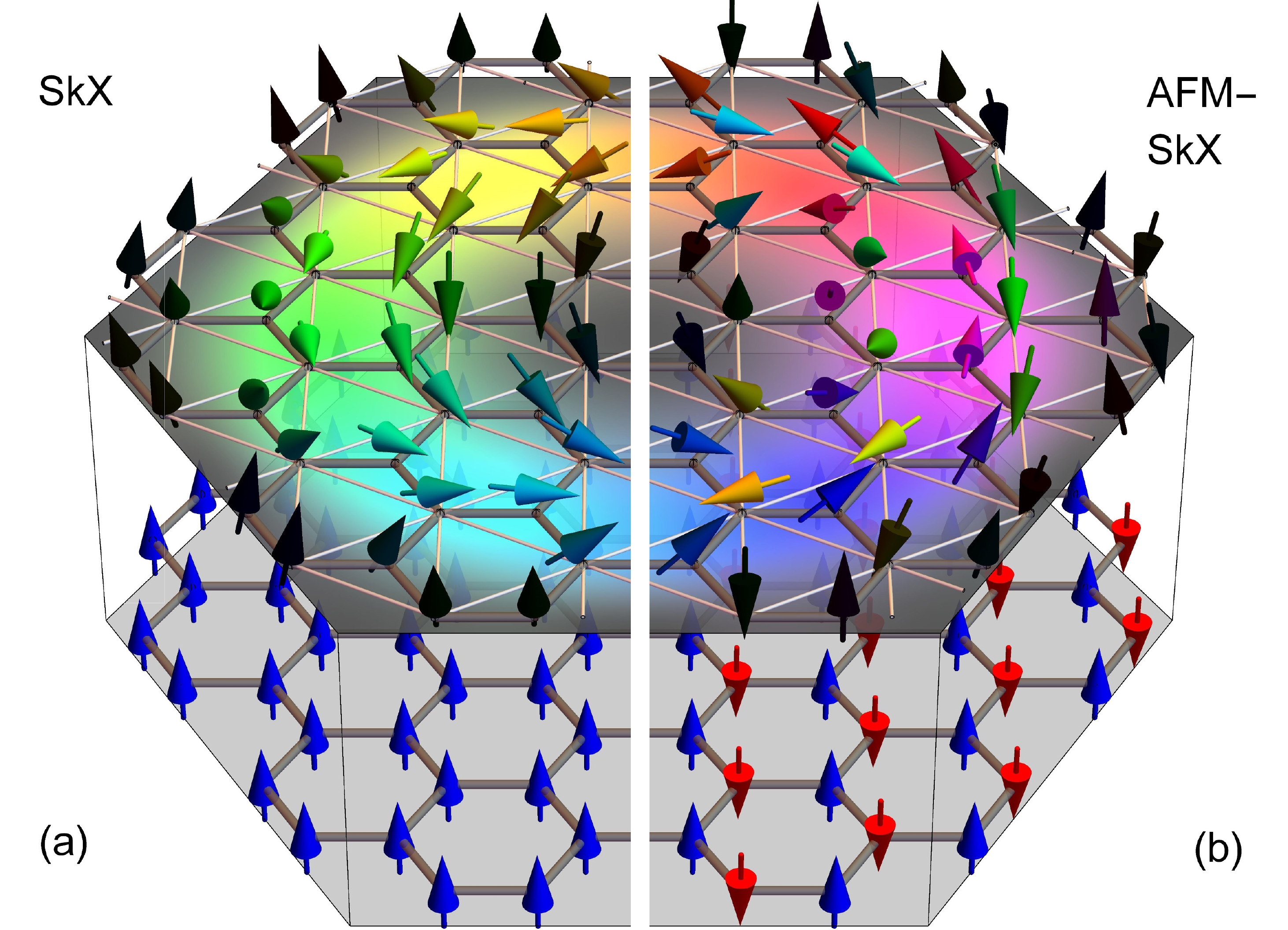}
  \caption{(a) Skyrmion and (b) antiferromagnetic skyrmion crystal on a honeycomb lattice. The spins at each site are represented by arrows. The lower hexagon represents (a) a ferromagnet and (b) a collinear antiferromagnet, on which the (antiferromagnetic) skyrmion layer has been deposited. Gray lines, forming the honeycomb lattice, represent exchange interactions with constant $J_1^{\sublattice{AB}}$; see text. White, thin lines visualize the exchange coupling within a sublattice (among second-nearest sites) $J_1$.}
  \label{fig:skyrmion_cell}
\end{figure}

In this \changeb{Rapid Communication}, \change{we predict the generation of stable AFM-SkXs by coupling a bipartite skyrmion material to a collinear antiferromagnetic layer (Fig.~\ref{fig:skyrmion_cell}b). 
The interlayer interaction acts as staggered magnetic field, which flip\changeb{s} the spin\changeb{s} of the SkX on one sublattice. 
The approach is generally applicable, as it can turn \emph{every} established phase of conventional SkXs into an AFM-SkX phase, irrespective of the skyrmion-generating mechanism. As an example, we apply the method to frustrated spins on a honeycomb lattice, i.\,e., two triangular sublattices that exhibit SkXs via frustrated exchange interactions (cf.\ Ref.~\onlinecite{okubo2012multiple}).}

If both sublattices of the AFM-SkX are equivalent, there is no THE\@. However, we find a topological spin Hall effect (TSHE). \change{Since the TSHE arises in a single two-dimensional layer, it is clearly distinguished from that in the surrogate system discussed in Refs.~\onlinecite{zhang2016magnetic,buhl2017topological}}. For inequivalent sublattices the THE becomes also nonzero, \change{which \changeb{may become} considerable for applications, once the predicted existence of AFM-SkXs has been realized experimentally.}

\paragraph*{\change{Generation of} AFM skyrmion crystals.}
First, we \change{present our approach to create a stable AFM-SkX starting from a known SkX phase. We take two copies of that two-dimensional system and couple them to a collinear antiferromagnet. This inverts the spins of one sublattice and yields a stable AFM-SkX with the parameters of the initial SkX\@. This approach is generally applicable, as it does not depend on the SkX-generating mechanism.}

\change{As an example we take} a honeycomb lattice \change{featuring two} triangular sublattices, \sublattice{A} and \sublattice{B}, \change{which both exhibit} a SkX \change{produced by frustration (Ref.~\onlinecite{okubo2012multiple}). The sublattice skyrmions are} stabilized by an external magnetic field and by thermal fluctuations; \change{they can be understood as superposition of three degenerate spin spirals which form the ground state for zero temperature and no magnetic field. To make the sublattice SkXs match we} add a weak inter-sublattice coupling \change{(results for a realistic inter-sublattice coupling are shown below).} 

The system is described by the Hamiltonian~\cite{okubo2012multiple}
\begin{align}
H_\mathrm{MC} =-\frac{1}{2}\sum_{i,j}J_{ij}\vec{s}_i \cdot \vec{s}_j - \sum_is_i^z B_i,\label{eq:HamiltonianMC}
\end{align}
in which $J_{ij}$ are Heisenberg exchange constants ($i$ and $j$ site indices). We take into account nearest- ($J_1$) and third-nearest-neighbor exchange ($J_3$) within each sublattice and add inter-sublattice coupling ($J_1^{\sublattice{AB}}$). The Zeeman term provides coupling to the external magnetic field $B_i$ along the $z$ direction. All energies are given in units of a global constant $J$. The magnetic configuration $\{\vec{s}_{i}\}$ is computed by classical Monte Carlo simulations ($\vec{s}_{i}$ spin of unit length).

\begin{figure*}
  \centering
   \includegraphics[width=\textwidth]{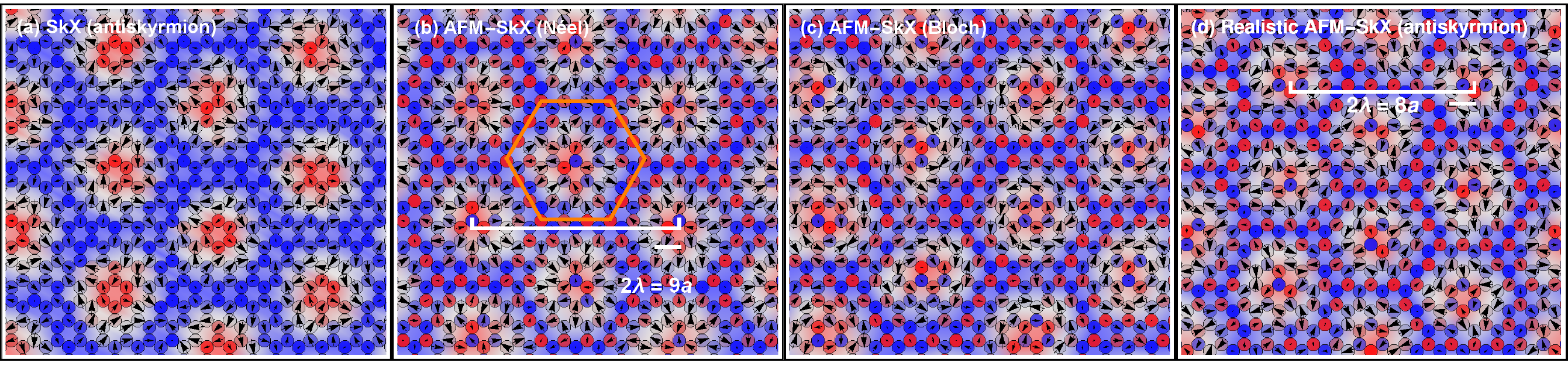}
  \caption{Antiferromagnetic skyrmion crystals on a honeycomb lattice of size $\lambda = 4.5\,a$ [see (b); $a$ lattice constant] characterized by vorticity $v$ and helicity $\gamma$ (azimuth of a spin $\Phi=v\phi+\gamma$, where $\phi$ is the azimuth of the position vector with respect to the skyrmions center). \change{Blue (red) circles denote a positive (negative) $z$-component of the spins, whereas arrows represent their in-plane components (modulus and direction).} (a) Crystal of antiskyrmions ($v=-1$; $\gamma\approx -\pi/3$). (b) Crystal of antiferromagnetic N{\'e}el-type skyrmions (sublattice skyrmions $v=+1$; $\gamma\approx \frac{\pi}{2}\pm\frac{\pi}{2}$). (c) Crystal of antiferromagnetic Bloch-type skyrmions (sublattice skyrmions $v=+1$; $\gamma\approx \pi\pm\frac{\pi}{2}$). Parameters: $J_1 = 1.63042\,J$, $J_3 = -J$, and $k_{\mathrm{B}} T = 2.5\,J$; for (a) $J_1^{\sublattice{AB}} = +0.05\,J$ and $B_\sublattice{A} = +B_\sublattice{B} = 0.9\,J$, while for (b) and (c) $J_1^{\sublattice{AB}} = -0.05\,J$ and $B_\sublattice{A} = -B_\sublattice{B} = 0.9\,J$. \change{This corresponds to a coupling to a collinear antiferromagnet with strength $0.9\,J$}. Cluster size: $36 \times 36$ sites per sublattice. \change{(d) Realistic antiferromagnetic antiskyrmion crystal. Parameters: $J_1 = 0.42\,J$, $J_3 = -0.665\,J$, $J_1^{\sublattice{AB}} = -1.751\,J$ on a $32 \times 32$ cluster; $T$ and $B$ as in (b) and (c). The size of the antiferromagnetic skyrmions is reduced ($\lambda = 4\,a$).}}
  \label{fig:state_sk_afm_sk}
\end{figure*}

As a prerequisite, \change{a weak} inter-sublattice coupling $J_1^{\sublattice{AB}} \ll J$ (see caption of Fig.~\ref{fig:skyrmion_cell}) ensures that the skyrmion-center locations of \sublattice{A} and \sublattice{B} adjust to each other. \change{This way} the lattice constant of the SkX and the magnetic phase diagram remain almost unchanged (with respect to the uncoupled SkXs~\cite{okubo2012multiple}). An exemplary result for a conventional SkX is shown in Fig.~\ref{fig:state_sk_afm_sk}a. 

\change{To create an AFM-SkX the spins of one sublattice have to be reversed. This would require an unrealistic staggered magnetic field $B_\sublattice{A} = -B_\sublattice{B}$. Instead we mimic it} by placing the skyrmion lattice on a \change{collinear antiferromagnet} with strong out-of-plane uniaxial anisotropy (Fig.~\ref{fig:skyrmion_cell}b). \change{For matching sublattices the} inter-sublattice coupling $J_1^{\sublattice{AB}}$ \change{has to be chosen negative.} The resulting AFM-SkXs on top of an antiferromagnet (Fig.~\ref{fig:skyrmion_cell}b) have the same energy and exhibit the same geometry (compare Fig.~\ref{fig:state_sk_afm_sk}a with b and c) as the SkXs (Fig.~\ref{fig:skyrmion_cell}a).  

\change{Special properties of the SkXs attributed to frustration survive our approach:} helicity (i.\,e.,  N{\'e}el- or Bloch-type skyrmions), winding (i.\,e., skyrmions or antiskyrmions with topological charge $\mp 1$), and skyrmion-center locations are not fixed for both SkXs and AFM-SkXs (Fig.~\ref{fig:state_sk_afm_sk}).

\change{In real materials the sublattices \sublattice{A} and \sublattice{B} are strongly coupled; $J_1^{\sublattice{AB}}\gg0$. Nevertheless, our simulations show that AFM-SkXs can still be stabilized (Fig.~\ref{fig:state_sk_afm_sk}d) but  lattice constant, stabilizing field, and temperature of the initial sublattice skyrmions can not be carried over to the resulting AFM-SkX\@.} 

A synopsis: AFM-SkXs can be produced by coupling a two-sublattice SkX to an antiferromagnetic layer (Fig.~\ref{fig:skyrmion_cell}b). This approach is valid irrespective of the physical mechanism that stabilizes the SkX (frustration~\cite{okubo2012multiple}, Dzyaloshinskii-Moriya interaction~\cite{do2009skyrmions} or anisotropy~\cite{leonov2015multiply}). The novel AFM-SkX \change{state motivates} to calculate the THE and TSHE\@.

\paragraph*{Electron transport in (AFM) skyrmion textures.}
In a tight-binding model the interaction of electrons with a (AFM) skyrmion texture $\{ \vec{s}_{i} \}$ is described by the Hamiltonian~\cite{hamamoto2015quantized} 
\begin{align} 
  H & = \sum_{ij} t_{ij} \,c_{i}^\dagger c_{j} + m \sum_{i} \vec{s}_{i} \cdot (c_{i}^\dagger \vec{\sigma}c_{i}),
  \label{eq:ham_the} 
\end{align}
($c_{i}^\dagger$ and $c_{i}$ creation and annihilation operators, respectively; $\vec{\sigma}$ vector of Pauli matrices). The hopping from site $i$ to site $j$ is quantified by $t_{ij}$, the coupling to the skyrmion texture by $m$.

The transverse charge conductivity $\sigma_{xy}$ at the Fermi energy $E_{\mathrm{F}}$ is calculated from the Kubo formula~\cite{nagaosa2010anomalous}
\begin{align}
  \sigma_{xy}(E_\mathrm{F}) & = \frac{e^{2}}{h} \frac{1}{2\pi} \sum_{n} \int_{\mathrm{BZ}} \Omega_{n}^{z}(\vec{k}) \, f(E_{n}(\vec{k}) - E_\mathrm{F}) \,\mathrm{d}^{2}k\label{eq:kubo}
\end{align}
[BZ Brillouin zone; $\vec{k} = (k_{x}, k_{y})$]. The sum runs over all bands $n$. $f(E)$ is the Fermi distribution function at temperature $T$; $e$, $h$, and $k_{\mathrm{B}}$ denote the electron charge, the Planck constant, and the Boltzmann constant, respectively. The Berry curvature \changeb{(a general version that also describes spin transport)},
\begin{align*}
  \vec{\Omega}_n(\vec{k}) & = \mathrm{i} \sum_{m \ne n}
  \frac{\braket{u_{n\vec{k}} | \nabla_{\vec{k}}\mathcal{M} H_{\vec{k}}| u_{m\vec{k}}} \times\braket{u_{m\vec{k}}|\nabla_{\vec{k}} H_{\vec{k}} | u_{n\vec{k}}}}{(E_{n\vec{k}} - E_{m \vec{k}})^2},
\end{align*}
is determined from the eigenvectors $u_{n}(\vec{k})$ with eigenenergies $E_{n \vec{k}}$ of the $\vec{k}$-dependent Hamiltonian $H_{\vec{k}}$~\cite{gradhand2012first}. For the topological Hall conductivity $\sigma_{xy}$ in skyrmion textures, the $(2\, n)\times (2\, n)$ matrix $\mathcal{M}$ is a unit matrix. If $E_\mathrm{F}$ lies within the band gap above the $n$th band, $\sigma_{xy}$ is proportional to the winding number~\cite{Hatsugai1993,Hatsugai1993a}, $w_{n} = \sum_{m \le n} C_m$, that is the accumulation of the integer Chern numbers $C_{m} = \frac{1}{2\pi} \int_{\mathrm{BZ}}\Omega_{m}^{(z)}(\vec{k}) \,\mathrm{d}^{2}k$.

For the spin conductivity, $\mathcal{M} = \operatorname{diag}(\vec{s}_1 \cdot \vec{\sigma},\dots,\vec{s}_n \cdot \vec{\sigma})$ accounts for the alignment of the electron spin with the skyrmion texture. Additionally, Eq.~\eqref{eq:kubo} has to be multiplied by $\hbar/(2e)$ to reflect spin instead of charge transport. For the spin conductivity in AFM-SkXs the sign of the entries are reversed for the sublattice with negative net magnetization, since a locally parallel aligned spin means spin up or down in the respective sublattice.

In the following, we utilize skyrmion textures on the honeycomb lattice that enter Eq.~\eqref{eq:ham_the} by superposing three spin spirals, as in Ref.~\onlinecite{okubo2012multiple} (Fig.~\ref{fig:skyrmion_cell}a). An AFM-SkX is then constructed by reversing the spins in one of the sublattices (Fig.~\ref{fig:skyrmion_cell}b). These textures are idealized versions of those generated from $H_\mathrm{MC}$ (Fig.~\ref{fig:state_sk_afm_sk}).

\begin{figure}
  \centering
  \includegraphics[width=\columnwidth]{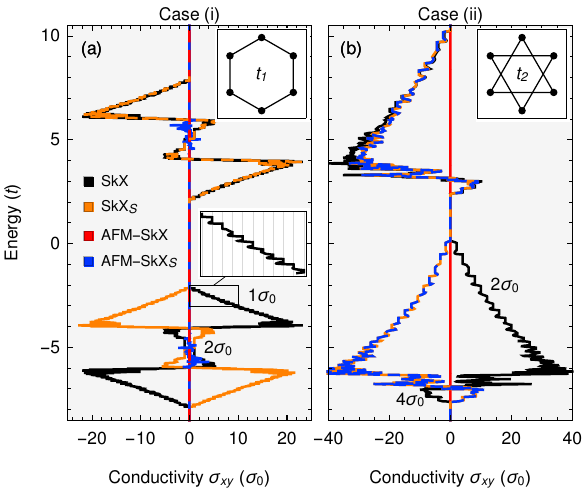}
  \caption{Topological Hall conductivities in skyrmion crystals [black: charge conductivity (SkX); orange: spin conductivity (SkX$_S$)] and in antiferromagnetic skyrmion crystals [red: charge conductivity (AFM-SkX); blue: spin conductivity (AFM-SkX$_S$)] \change{with 72 sites in the unit cell}. The tight-binding parameters read $t_1 = t$, $t_2 = 0$ [a, case (i)] and $t_1 = 0$, $t_2 = t$ [b, case (ii)]; the coupling to the skyrmion texture equals $m = 5 \, t$. The hopping strengths are sketched in the insets. Conductivities are quantized in units of $\sigma_0 = e^2 / h$ (charge) and $\sigma_0 = e / (4\pi)$ (spin); cf. Refs.~\onlinecite{gobel2017THEskyrmion,gobel2017QHE}.}
  \label{fig:cond_tria_honeycomb}
\end{figure}

\paragraph*{Topological Hall effects in skyrmion crystals.} For the topological Hall effect (THE) in a SkX (Fig.~\ref{fig:skyrmion_cell}a), we consider two generic cases: (i) nearest-neighbor hopping strength $t_1 = t$ and second-nearest neighbor hopping strength $t_2 = 0$ as well as (ii) $t_1 = 0$ and $t_2 = t$ (cf.\ the insets in Fig.~\ref{fig:cond_tria_honeycomb}).

For large coupling $m$ to the skyrmion texture ($m = 5 \, t$ in Fig.~\ref{fig:cond_tria_honeycomb}), the band structure is energetically split into two blocks (rigidly shifted by $\pm m$). In each of the blocks, the electron spin is aligned parallel (lower block) or antiparallel (upper block) to the texture. As a result, the respective energy-dependent \change{transverse} conductivities have opposite sign and exhibit (almost) identical shapes~\cite{gobel2017THEskyrmion}.

The above qualitative picture is nicely reproduced by the computed THE of case (i) (black line in Fig.~\ref{fig:cond_tria_honeycomb}a; cf.\ Ref.~\onlinecite{gobel2017QHE}). Within each block, the conductivity is antisymmetric because the sublattices are equivalent. The bands of the lower (upper) block carry Chern number $-1$ ($+1$), except for bands close to a van Hove singularity of the zero-field band structure (at $\pm m \pm t$), as is explained in Refs.~\onlinecite{gobel2017THEskyrmion} and~\onlinecite{gobel2017QHE} \change{(at the associated energies the Fermi lines change their character from electron to hole pockets)}. The latter bands compensate the accumulated large Chern numbers of all other bands in their block \change{and bring about a sign change in $\sigma_{xy}$}. 

For case (ii) (two uncoupled triangular sublattices~\cite{gobel2017THEskyrmion}), we find the separation into two blocks as well. Every band is almost degenerate (minimal splitting due to $E(\vec{k})\ne E(-\vec{k})$ for bands of both sublattices; see Fig.~S2 in the Supplemental Material~\footnote{See Supplemental Material at [URL will be inserted by publisher] for band structures of the SkX and AFM-SkX}). Thus, the conductivity shows steps in units of $2 \cdot e^2 / h$ (Fig.~\ref{fig:cond_tria_honeycomb}b), that is twice as large as in case (i). 

The alignment of the spins (parallel or antiparallel) with the skyrmion texture results in a transverse spin-polarized current~\cite{yin2015topological}. The magnitude of the spin conductivity corresponds to the charge conductivity (in the block-separated case for large coupling $m$). Spin and charge current are inseparable.

\paragraph*{Topological spin Hall effect in AFM skyrmion crystals.} We proceed with the generic cases for the AFM-SkXs (Fig.~\ref{fig:skyrmion_cell}b). Case (i) exhibits no considerable transverse transport because the emergent field fluctuates around zero, yielding zero net field. In case (ii) the topological Hall conductivity is zero as well; this is explained by the two sublattices having opposite emergent fields. However, we find a topological spin Hall effect.

\begin{figure}
  \centering
  \includegraphics[width=\columnwidth]{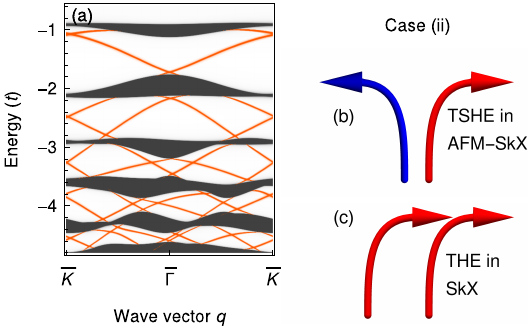}
  \caption{Topological spin Hall effect in an antiferromagnetic
skyrmion crystal of generic case (ii) with $32$ sites in the unit cell. (a) Electronic structure at the edge of the semi-infinite sample computed by Green function renormalization~\cite{henk1993subroutine,bodicker1994interface}. Black: bulk states, orange: edge states. (b) Deflection of electrons with opposite spins (blue and red arrows) in an AFM-SkX (schematic). (c) Deflection of electrons with equal spins in a SkX.}
  \label{fig:TSHE}
\end{figure}

The bands of case (ii) are two-fold degenerate because the sublattices are equivalent [$E^{\sublattice{A}}(\vec{k})=E^{\sublattice{B}}(\vec{k})$]. The spin is aligned parallel (lower block) or antiparallel (upper block) to the texture of the respective sublattice. Since the sublattices are decoupled ($t_1 = 0$), the electrons are localized exclusively in either sublattice. This causes a spin-up current (from the sublattice with positive net magnetization) \emph{and} a spin-down current (from the other sublattice, with negative net magnetization). Hence, a TSHE occurs which is identical to the (spin-polarized) THE in the SkX (Fig.~\ref{fig:TSHE}b,c; cf.\ the blue and orange lines in Fig.~\ref{fig:cond_tria_honeycomb}b). For the AFM-SkX we find a pure spin current; the THE is zero.

In each of the bulk-band gaps the number of right--propagating edge states is identical to that of left-propagating ones (Fig.~\ref{fig:TSHE}a): there is no charge transport, i.\,e., no THE\@. Since the edge states `live' on different sublattices they carry opposite spin because their spins are aligned with the associated sublattice texture. The emergent fields of the individual sublattices have opposite signs; thus, they deflect electrons of \emph{opposite spin into opposite directions} (Fig.~\ref{fig:TSHE}b). The result is a TSHE\@. Recall that in a SkX the identical emergent fields of the sublattices  deflect electrons of the \emph{same spin into the same direction} (Fig.~\ref{fig:TSHE}c); hence the spin conductivities for AFM-SkX and SkX are identical, but in the AFM-SkX there is no effective transverse charge current.

For intermediate and more general cases, i.\,e., $t_1\ne 0$ and $t_2\ne 0$, the results lie between case (i) and (ii) (see Fig.~S1 in the Supplemental Material~\footnote{See Supplemental Material at [URL will be inserted by publisher] for topological (spin) Hall conductivities for the case $t_1  = t_2 = t/2$.}). The TSHE in an AFM-SkX is nonzero as long as $t_2>0$. The THE is zero in any case.

Summarizing, one finds a THE of spin-polarized electrons in SkXs (Fig.~\ref{fig:skyrmion_cell}a) and a TSHE in AFM-SkXs (Fig.~\ref{fig:skyrmion_cell}b)\change{, that is, the analogues to Hall and spin Hall physics in a single two-dimensional layer, as distinguished from the surrogate multilayer system of Refs.~\cite{zhang2016magnetic,buhl2017topological}.}

\begin{figure}
  \centering
  \includegraphics[width=\columnwidth]{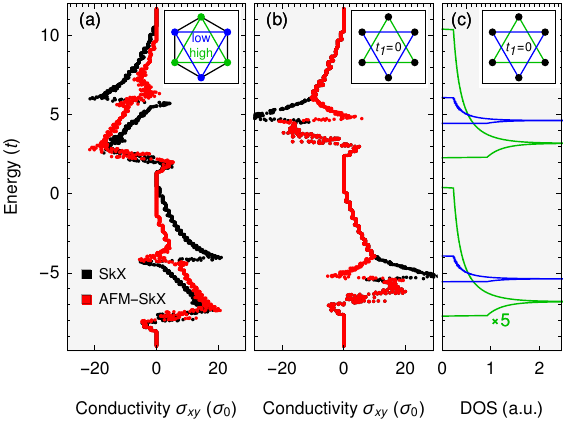}
  \caption{Topological Hall conductivity (SkX: black; asymmetric AFM-SkX: red. In units of $\sigma_{0} = e^2 / h$). (a) Conductivity $\sigma_{xy}$ versus energy for differing on-site energies and second-nearest neighbor strengths: $t_1 = 0.75 \, t$, $t_2^\sublattice{A} = 0.2 \, t \ne t_2^\sublattice{B} = t$, $\delta \epsilon = 2t$. (b) As (a) but with $\delta \epsilon = 0$ and $t_1 = 0$. The coupling to the skyrmion texture is $m = 5 \, t$ in all cases. (c) DOS of the zero-field band structure for the parameters of panel~(b).}
  \label{fig:real}
\end{figure}

\paragraph*{Topological Hall effect in asymmetric AFM skyrmion crystals.} Having discussed generic cases, we proceed with sublattice-asymmetric AFM-SkXs (e.g., in crystals consisting of two different elements), which is modeled by setting $t_2^\sublattice{A} \ne t_2^\sublattice{B}$ and by differing on-site energies, $\delta\epsilon=\epsilon^{\sublattice{A}} - \epsilon^{\sublattice{B}}\ne 0$. The topological Hall conductivity exhibits the band-block separation (Figs.~\ref{fig:real}a and b) and is nonzero in any case. 

To clarify these findings we consider the tight-binding Hamiltonian~\eqref{eq:ham_the} without spin texture ($m=0$), with parameters as in Fig.~\ref{fig:real}b (uncoupled sublattices). The density of states (DOS) of the resulting two bands (one band per sublattice) is shown in Fig.~\ref{fig:real}c. Comparing SkX and AFM-SkX, the sublattice skyrmions on \sublattice{A} (green curve) have the same winding, while for \sublattice{B} (blue) they have opposite winding. Therefore, in regions in which the two zero-field bands (green and blue) do not overlap in energy, the topological Hall conductivities of a SkX and an AFM-SkX are identical. 

The contribution of the narrow band (blue) has to be subtracted (added) from (to) the conductivity corresponding to the green band for the AFM-SkX (SkX) because of the opposite (identical) winding of the sublattice skyrmion (see Ref.~\onlinecite{gobel2017QHE}).

For nonzero $t_{1}$ and $\delta \epsilon$ (Fig.~\ref{fig:real}a) a sublattice separation of the bands is no longer given, but the conductivity does not change qualitatively. It is even possible that the topological Hall conductivity of an AFM-SkX exceeds that of a SkX (see Fig.~S3 in the Supplemental Material~\footnote{See Supplemental Material at [URL will be inserted by publisher] for topological Hall conductivities for the case $t_2^\sublattice{A}  = -t_2^\sublattice{B}$.}).

\paragraph*{\change{Conclusion}.} \change{In this \changeb{Rapid Communication}, we predict the generation of stable antiferromagnetic skyrmion crystals. These} systems can in principle be realized on \textit{any} bipartite lattice --- provided the individual sublattices exhibit a conventional skyrmion crystal \change{(irrespective of the generating mechanism)} --- by growing it onto a collinear antiferromagnet (Fig.~\ref{fig:skyrmion_cell}b). 

For equivalent sublattices, there is no topological Hall effect but a topological spin Hall effect. Furthermore, asymmetric antiferromagnetic skyrmion crystals (i.\,e., with inequivalent sublattices) exhibit a topological Hall effect. These findings are valid also for metastable single antiferromagnetic skyrmions; see Refs.~\onlinecite{zhang2016antiferromagnetic,barker2016static,PhysRevB.95.054421}. Very recently ferrimagnetic skyrmions have been found in GdFeCo films~\cite{woo2017current}. The magnetic moments of the two sublattices are inequivalent and a topological Hall effect is measurable, which corroborates our analysis.


Besides the potential of \textit{stable} AFM-SkX\change{s} for applications, the Hamiltonian $H_\mathrm{MC}$ of Eq.~\eqref{eq:HamiltonianMC} motivates further theoretical investigations. An example is transport via magnons, studied in stable magnetic configurations. One may compare the topological magnon Hall effects in skyrmion crystals~\cite{mook2017magnon} with that in antiferromagnetic skyrmion crystals.

\begin{acknowledgments}
This work is supported by Priority Program SPP~1666 of Deutsche Forschungsgemeinschaft (DFG).
\end{acknowledgments}

\bibliography{short,MyLibrary}

\begin{thebibliography}{49}
\expandafter\ifx\csname natexlab\endcsname\relax\def\natexlab#1{#1}\fi
\expandafter\ifx\csname bibnamefont\endcsname\relax
  \def\bibnamefont#1{#1}\fi
\expandafter\ifx\csname bibfnamefont\endcsname\relax
  \def\bibfnamefont#1{#1}\fi
\expandafter\ifx\csname citenamefont\endcsname\relax
  \def\citenamefont#1{#1}\fi
\expandafter\ifx\csname url\endcsname\relax
  \def\url#1{\texttt{#1}}\fi
\expandafter\ifx\csname urlprefix\endcsname\relax\def\urlprefix{URL }\fi
\providecommand{\bibinfo}[2]{#2}
\providecommand{\eprint}[2][]{\url{#2}}

\bibitem[{\citenamefont{Skyrme}(1962)}]{skyrme1962unified}
\bibinfo{author}{\bibfnamefont{T.~H.~R.} \bibnamefont{Skyrme}},
  \bibinfo{journal}{Nuclear Physics} \textbf{\bibinfo{volume}{31}},
  \bibinfo{pages}{556} (\bibinfo{year}{1962}).

\bibitem[{\citenamefont{Bogdanov and
  Yablonskii}(1989)}]{bogdanov1989thermodynamically}
\bibinfo{author}{\bibfnamefont{A.}~\bibnamefont{Bogdanov}} \bibnamefont{and}
  \bibinfo{author}{\bibfnamefont{D.}~\bibnamefont{Yablonskii}},
  \bibinfo{journal}{Zh. Eksp. Teor. Fiz} \textbf{\bibinfo{volume}{95}},
  \bibinfo{pages}{182} (\bibinfo{year}{1989}).

\bibitem[{\citenamefont{Bogdanov and
  Hubert}(1994)}]{bogdanov1994thermodynamically}
\bibinfo{author}{\bibfnamefont{A.}~\bibnamefont{Bogdanov}} \bibnamefont{and}
  \bibinfo{author}{\bibfnamefont{A.}~\bibnamefont{Hubert}},
  \bibinfo{journal}{J. Magn.\ Magn.\ Mater.} \textbf{\bibinfo{volume}{138}},
  \bibinfo{pages}{255} (\bibinfo{year}{1994}).

\bibitem[{\citenamefont{R{\"o}{\ss}ler
  et~al.}(2006)\citenamefont{R{\"o}{\ss}ler, Bogdanov, and
  Pfleiderer}}]{rossler2006spontaneous}
\bibinfo{author}{\bibfnamefont{U.}~\bibnamefont{R{\"o}{\ss}ler}},
  \bibinfo{author}{\bibfnamefont{A.}~\bibnamefont{Bogdanov}}, \bibnamefont{and}
  \bibinfo{author}{\bibfnamefont{C.}~\bibnamefont{Pfleiderer}},
  \bibinfo{journal}{Nature} \textbf{\bibinfo{volume}{442}},
  \bibinfo{pages}{797} (\bibinfo{year}{2006}).

\bibitem[{\citenamefont{M{\"u}hlbauer et~al.}(2009)\citenamefont{M{\"u}hlbauer,
  Binz, Jonietz, Pfleiderer, Rosch, Neubauer, Georgii, and
  B{\"o}ni}}]{muhlbauer2009skyrmion}
\bibinfo{author}{\bibfnamefont{S.}~\bibnamefont{M{\"u}hlbauer}},
  \bibinfo{author}{\bibfnamefont{B.}~\bibnamefont{Binz}},
  \bibinfo{author}{\bibfnamefont{F.}~\bibnamefont{Jonietz}},
  \bibinfo{author}{\bibfnamefont{C.}~\bibnamefont{Pfleiderer}},
  \bibinfo{author}{\bibfnamefont{A.}~\bibnamefont{Rosch}},
  \bibinfo{author}{\bibfnamefont{A.}~\bibnamefont{Neubauer}},
  \bibinfo{author}{\bibfnamefont{R.}~\bibnamefont{Georgii}}, \bibnamefont{and}
  \bibinfo{author}{\bibfnamefont{P.}~\bibnamefont{B{\"o}ni}},
  \bibinfo{journal}{Science} \textbf{\bibinfo{volume}{323}},
  \bibinfo{pages}{915} (\bibinfo{year}{2009}).

\bibitem[{\citenamefont{Dzyaloshinsky}(1958)}]{dzyaloshinsky1958thermodynamic}
\bibinfo{author}{\bibfnamefont{I.}~\bibnamefont{Dzyaloshinsky}},
  \bibinfo{journal}{J. Phys.\ Chem.\ Sol.} \textbf{\bibinfo{volume}{4}},
  \bibinfo{pages}{241} (\bibinfo{year}{1958}).

\bibitem[{\citenamefont{Moriya}(1960)}]{moriya1960anisotropic}
\bibinfo{author}{\bibfnamefont{T.}~\bibnamefont{Moriya}},
  \bibinfo{journal}{Phys.\ Rev.} \textbf{\bibinfo{volume}{120}},
  \bibinfo{pages}{91} (\bibinfo{year}{1960}).

\bibitem[{\citenamefont{Nagaosa and Tokura}(2013)}]{nagaosa2013topological}
\bibinfo{author}{\bibfnamefont{N.}~\bibnamefont{Nagaosa}} \bibnamefont{and}
  \bibinfo{author}{\bibfnamefont{Y.}~\bibnamefont{Tokura}},
  \bibinfo{journal}{Nature Nanotechnology} \textbf{\bibinfo{volume}{8}},
  \bibinfo{pages}{899} (\bibinfo{year}{2013}).

\bibitem[{\citenamefont{Okubo et~al.}(2012)\citenamefont{Okubo, Chung, and
  Kawamura}}]{okubo2012multiple}
\bibinfo{author}{\bibfnamefont{T.}~\bibnamefont{Okubo}},
  \bibinfo{author}{\bibfnamefont{S.}~\bibnamefont{Chung}}, \bibnamefont{and}
  \bibinfo{author}{\bibfnamefont{H.}~\bibnamefont{Kawamura}},
  \bibinfo{journal}{Phys.\ Rev.\ Lett.} \textbf{\bibinfo{volume}{108}},
  \bibinfo{pages}{017206} (\bibinfo{year}{2012}).

\bibitem[{\citenamefont{Fert et~al.}(2013)\citenamefont{Fert, Cros, and
  Sampaio}}]{fert2013skyrmions}
\bibinfo{author}{\bibfnamefont{A.}~\bibnamefont{Fert}},
  \bibinfo{author}{\bibfnamefont{V.}~\bibnamefont{Cros}}, \bibnamefont{and}
  \bibinfo{author}{\bibfnamefont{J.}~\bibnamefont{Sampaio}},
  \bibinfo{journal}{Nature Nanotechnol.} \textbf{\bibinfo{volume}{8}},
  \bibinfo{pages}{152} (\bibinfo{year}{2013}).

\bibitem[{\citenamefont{Wiesendanger}(2016)}]{wiesendanger2016nanoscale}
\bibinfo{author}{\bibfnamefont{R.}~\bibnamefont{Wiesendanger}},
  \bibinfo{journal}{Nature Reviews Materials} \textbf{\bibinfo{volume}{1}},
  \bibinfo{pages}{16044} (\bibinfo{year}{2016}).

\bibitem[{\citenamefont{Romming et~al.}(2013)\citenamefont{Romming, Hanneken,
  Menzel, Bickel, Wolter, von Bergmann, Kubetzka, and
  Wiesendanger}}]{romming2013writing}
\bibinfo{author}{\bibfnamefont{N.}~\bibnamefont{Romming}},
  \bibinfo{author}{\bibfnamefont{C.}~\bibnamefont{Hanneken}},
  \bibinfo{author}{\bibfnamefont{M.}~\bibnamefont{Menzel}},
  \bibinfo{author}{\bibfnamefont{J.~E.} \bibnamefont{Bickel}},
  \bibinfo{author}{\bibfnamefont{B.}~\bibnamefont{Wolter}},
  \bibinfo{author}{\bibfnamefont{K.}~\bibnamefont{von Bergmann}},
  \bibinfo{author}{\bibfnamefont{A.}~\bibnamefont{Kubetzka}}, \bibnamefont{and}
  \bibinfo{author}{\bibfnamefont{R.}~\bibnamefont{Wiesendanger}},
  \bibinfo{journal}{Science} \textbf{\bibinfo{volume}{341}},
  \bibinfo{pages}{636} (\bibinfo{year}{2013}).

\bibitem[{\citenamefont{Hsu et~al.}(2017)\citenamefont{Hsu, Kubetzka, Finco,
  Romming, von Bergmann, and Wiesendanger}}]{hsu2016electric}
\bibinfo{author}{\bibfnamefont{P.-J.} \bibnamefont{Hsu}},
  \bibinfo{author}{\bibfnamefont{A.}~\bibnamefont{Kubetzka}},
  \bibinfo{author}{\bibfnamefont{A.}~\bibnamefont{Finco}},
  \bibinfo{author}{\bibfnamefont{N.}~\bibnamefont{Romming}},
  \bibinfo{author}{\bibfnamefont{K.}~\bibnamefont{von Bergmann}},
  \bibnamefont{and}
  \bibinfo{author}{\bibfnamefont{R.}~\bibnamefont{Wiesendanger}},
  \bibinfo{journal}{Nature Nanotechnol.} \textbf{\bibinfo{volume}{12}},
  \bibinfo{pages}{123} (\bibinfo{year}{2017}).

\bibitem[{\citenamefont{Zhang et~al.}(2015{\natexlab{a}})\citenamefont{Zhang,
  Ezawa, and Zhou}}]{zhang2015magnetic}
\bibinfo{author}{\bibfnamefont{X.}~\bibnamefont{Zhang}},
  \bibinfo{author}{\bibfnamefont{M.}~\bibnamefont{Ezawa}}, \bibnamefont{and}
  \bibinfo{author}{\bibfnamefont{Y.}~\bibnamefont{Zhou}},
  \bibinfo{journal}{Scientific Reports} \textbf{\bibinfo{volume}{5}},
  \bibinfo{pages}{9400} (\bibinfo{year}{2015}{\natexlab{a}}).

\bibitem[{\citenamefont{Zhang et~al.}(2015{\natexlab{b}})\citenamefont{Zhang,
  Zhou, Ezawa, Zhao, and Zhao}}]{zhang2015magnetic2}
\bibinfo{author}{\bibfnamefont{X.}~\bibnamefont{Zhang}},
  \bibinfo{author}{\bibfnamefont{Y.}~\bibnamefont{Zhou}},
  \bibinfo{author}{\bibfnamefont{M.}~\bibnamefont{Ezawa}},
  \bibinfo{author}{\bibfnamefont{G.}~\bibnamefont{Zhao}}, \bibnamefont{and}
  \bibinfo{author}{\bibfnamefont{W.}~\bibnamefont{Zhao}},
  \bibinfo{journal}{Scientific Reports} \textbf{\bibinfo{volume}{5}},
  \bibinfo{pages}{11369} (\bibinfo{year}{2015}{\natexlab{b}}).

\bibitem[{\citenamefont{Jiang et~al.}(2015)\citenamefont{Jiang, Upadhyaya,
  Zhang, Yu, Jungfleisch, Fradin, Pearson, Tserkovnyak, Wang, Heinonen
  et~al.}}]{jiang2015blowing}
\bibinfo{author}{\bibfnamefont{W.}~\bibnamefont{Jiang}},
  \bibinfo{author}{\bibfnamefont{P.}~\bibnamefont{Upadhyaya}},
  \bibinfo{author}{\bibfnamefont{W.}~\bibnamefont{Zhang}},
  \bibinfo{author}{\bibfnamefont{G.}~\bibnamefont{Yu}},
  \bibinfo{author}{\bibfnamefont{M.~B.} \bibnamefont{Jungfleisch}},
  \bibinfo{author}{\bibfnamefont{F.~Y.} \bibnamefont{Fradin}},
  \bibinfo{author}{\bibfnamefont{J.~E.} \bibnamefont{Pearson}},
  \bibinfo{author}{\bibfnamefont{Y.}~\bibnamefont{Tserkovnyak}},
  \bibinfo{author}{\bibfnamefont{K.~L.} \bibnamefont{Wang}},
  \bibinfo{author}{\bibfnamefont{O.}~\bibnamefont{Heinonen}},
  \bibnamefont{et~al.}, \bibinfo{journal}{Science}
  \textbf{\bibinfo{volume}{349}}, \bibinfo{pages}{283} (\bibinfo{year}{2015}).

\bibitem[{\citenamefont{Boulle et~al.}(2016)\citenamefont{Boulle, Vogel, Yang,
  Pizzini, de~Souza~Chaves, Locatelli, Mente{\c{s}}, Sala, Buda-Prejbeanu,
  Klein et~al.}}]{boulle2016room}
\bibinfo{author}{\bibfnamefont{O.}~\bibnamefont{Boulle}},
  \bibinfo{author}{\bibfnamefont{J.}~\bibnamefont{Vogel}},
  \bibinfo{author}{\bibfnamefont{H.}~\bibnamefont{Yang}},
  \bibinfo{author}{\bibfnamefont{S.}~\bibnamefont{Pizzini}},
  \bibinfo{author}{\bibfnamefont{D.}~\bibnamefont{de~Souza~Chaves}},
  \bibinfo{author}{\bibfnamefont{A.}~\bibnamefont{Locatelli}},
  \bibinfo{author}{\bibfnamefont{T.~O.} \bibnamefont{Mente{\c{s}}}},
  \bibinfo{author}{\bibfnamefont{A.}~\bibnamefont{Sala}},
  \bibinfo{author}{\bibfnamefont{L.~D.} \bibnamefont{Buda-Prejbeanu}},
  \bibinfo{author}{\bibfnamefont{O.}~\bibnamefont{Klein}},
  \bibnamefont{et~al.}, \bibinfo{journal}{Nature Nanotechnol.}
  \textbf{\bibinfo{volume}{11}}, \bibinfo{pages}{449} (\bibinfo{year}{2016}).

\bibitem[{\citenamefont{Seki et~al.}(2012)\citenamefont{Seki, Yu, Ishiwata, and
  Tokura}}]{seki2012observation}
\bibinfo{author}{\bibfnamefont{S.}~\bibnamefont{Seki}},
  \bibinfo{author}{\bibfnamefont{X.}~\bibnamefont{Yu}},
  \bibinfo{author}{\bibfnamefont{S.}~\bibnamefont{Ishiwata}}, \bibnamefont{and}
  \bibinfo{author}{\bibfnamefont{Y.}~\bibnamefont{Tokura}},
  \bibinfo{journal}{Science} \textbf{\bibinfo{volume}{336}},
  \bibinfo{pages}{198} (\bibinfo{year}{2012}).

\bibitem[{\citenamefont{Woo et~al.}(2016)\citenamefont{Woo, Litzius,
  Kr{\"u}ger, Im, Caretta, Richter, Mann, Krone, Reeve, Weigand
  et~al.}}]{woo2016observation}
\bibinfo{author}{\bibfnamefont{S.}~\bibnamefont{Woo}},
  \bibinfo{author}{\bibfnamefont{K.}~\bibnamefont{Litzius}},
  \bibinfo{author}{\bibfnamefont{B.}~\bibnamefont{Kr{\"u}ger}},
  \bibinfo{author}{\bibfnamefont{M.-Y.} \bibnamefont{Im}},
  \bibinfo{author}{\bibfnamefont{L.}~\bibnamefont{Caretta}},
  \bibinfo{author}{\bibfnamefont{K.}~\bibnamefont{Richter}},
  \bibinfo{author}{\bibfnamefont{M.}~\bibnamefont{Mann}},
  \bibinfo{author}{\bibfnamefont{A.}~\bibnamefont{Krone}},
  \bibinfo{author}{\bibfnamefont{R.~M.} \bibnamefont{Reeve}},
  \bibinfo{author}{\bibfnamefont{M.}~\bibnamefont{Weigand}},
  \bibnamefont{et~al.}, \bibinfo{journal}{Nature Materials}
  \textbf{\bibinfo{volume}{15}}, \bibinfo{pages}{501} (\bibinfo{year}{2016}).

\bibitem[{\citenamefont{Neubauer et~al.}(2009)\citenamefont{Neubauer,
  Pfleiderer, Binz, Rosch, Ritz, Niklowitz, and
  B{\"o}ni}}]{neubauer2009topological}
\bibinfo{author}{\bibfnamefont{A.}~\bibnamefont{Neubauer}},
  \bibinfo{author}{\bibfnamefont{C.}~\bibnamefont{Pfleiderer}},
  \bibinfo{author}{\bibfnamefont{B.}~\bibnamefont{Binz}},
  \bibinfo{author}{\bibfnamefont{A.}~\bibnamefont{Rosch}},
  \bibinfo{author}{\bibfnamefont{R.}~\bibnamefont{Ritz}},
  \bibinfo{author}{\bibfnamefont{P.}~\bibnamefont{Niklowitz}},
  \bibnamefont{and} \bibinfo{author}{\bibfnamefont{P.}~\bibnamefont{B{\"o}ni}},
  \bibinfo{journal}{Phys.\ Rev.\ Lett.} \textbf{\bibinfo{volume}{102}},
  \bibinfo{pages}{186602} (\bibinfo{year}{2009}).

\bibitem[{\citenamefont{Schulz et~al.}(2012)\citenamefont{Schulz, Ritz, Bauer,
  Halder, Wagner, Franz, Pfleiderer, Everschor, Garst, and
  Rosch}}]{schulz2012emergent}
\bibinfo{author}{\bibfnamefont{T.}~\bibnamefont{Schulz}},
  \bibinfo{author}{\bibfnamefont{R.}~\bibnamefont{Ritz}},
  \bibinfo{author}{\bibfnamefont{A.}~\bibnamefont{Bauer}},
  \bibinfo{author}{\bibfnamefont{M.}~\bibnamefont{Halder}},
  \bibinfo{author}{\bibfnamefont{M.}~\bibnamefont{Wagner}},
  \bibinfo{author}{\bibfnamefont{C.}~\bibnamefont{Franz}},
  \bibinfo{author}{\bibfnamefont{C.}~\bibnamefont{Pfleiderer}},
  \bibinfo{author}{\bibfnamefont{K.}~\bibnamefont{Everschor}},
  \bibinfo{author}{\bibfnamefont{M.}~\bibnamefont{Garst}}, \bibnamefont{and}
  \bibinfo{author}{\bibfnamefont{A.}~\bibnamefont{Rosch}},
  \bibinfo{journal}{Nature Phys.} \textbf{\bibinfo{volume}{8}},
  \bibinfo{pages}{301} (\bibinfo{year}{2012}).

\bibitem[{\citenamefont{Kanazawa et~al.}(2011)\citenamefont{Kanazawa, Onose,
  Arima, Okuyama, Ohoyama, Wakimoto, Kakurai, Ishiwata, and
  Tokura}}]{kanazawa2011large}
\bibinfo{author}{\bibfnamefont{N.}~\bibnamefont{Kanazawa}},
  \bibinfo{author}{\bibfnamefont{Y.}~\bibnamefont{Onose}},
  \bibinfo{author}{\bibfnamefont{T.}~\bibnamefont{Arima}},
  \bibinfo{author}{\bibfnamefont{D.}~\bibnamefont{Okuyama}},
  \bibinfo{author}{\bibfnamefont{K.}~\bibnamefont{Ohoyama}},
  \bibinfo{author}{\bibfnamefont{S.}~\bibnamefont{Wakimoto}},
  \bibinfo{author}{\bibfnamefont{K.}~\bibnamefont{Kakurai}},
  \bibinfo{author}{\bibfnamefont{S.}~\bibnamefont{Ishiwata}}, \bibnamefont{and}
  \bibinfo{author}{\bibfnamefont{Y.}~\bibnamefont{Tokura}},
  \bibinfo{journal}{Phys.\ Rev.\ Lett.} \textbf{\bibinfo{volume}{106}},
  \bibinfo{pages}{156603} (\bibinfo{year}{2011}).

\bibitem[{\citenamefont{Lee et~al.}(2009)\citenamefont{Lee, Kang, Onose,
  Tokura, and Ong}}]{lee2009unusual}
\bibinfo{author}{\bibfnamefont{M.}~\bibnamefont{Lee}},
  \bibinfo{author}{\bibfnamefont{W.}~\bibnamefont{Kang}},
  \bibinfo{author}{\bibfnamefont{Y.}~\bibnamefont{Onose}},
  \bibinfo{author}{\bibfnamefont{Y.}~\bibnamefont{Tokura}}, \bibnamefont{and}
  \bibinfo{author}{\bibfnamefont{N.}~\bibnamefont{Ong}},
  \bibinfo{journal}{Phys.\ Rev.\ Lett.} \textbf{\bibinfo{volume}{102}},
  \bibinfo{pages}{186601} (\bibinfo{year}{2009}).

\bibitem[{\citenamefont{Li et~al.}(2013)\citenamefont{Li, Kanazawa, Yu,
  Tsukazaki, Kawasaki, Ichikawa, Jin, Kagawa, and Tokura}}]{li2013robust}
\bibinfo{author}{\bibfnamefont{Y.}~\bibnamefont{Li}},
  \bibinfo{author}{\bibfnamefont{N.}~\bibnamefont{Kanazawa}},
  \bibinfo{author}{\bibfnamefont{X.}~\bibnamefont{Yu}},
  \bibinfo{author}{\bibfnamefont{A.}~\bibnamefont{Tsukazaki}},
  \bibinfo{author}{\bibfnamefont{M.}~\bibnamefont{Kawasaki}},
  \bibinfo{author}{\bibfnamefont{M.}~\bibnamefont{Ichikawa}},
  \bibinfo{author}{\bibfnamefont{X.}~\bibnamefont{Jin}},
  \bibinfo{author}{\bibfnamefont{F.}~\bibnamefont{Kagawa}}, \bibnamefont{and}
  \bibinfo{author}{\bibfnamefont{Y.}~\bibnamefont{Tokura}},
  \bibinfo{journal}{Phys.\ Rev.\ Lett.} \textbf{\bibinfo{volume}{110}},
  \bibinfo{pages}{117202} (\bibinfo{year}{2013}).

\bibitem[{\citenamefont{Bruno et~al.}(2004)\citenamefont{Bruno, Dugaev, and
  Taillefumier}}]{bruno2004topological}
\bibinfo{author}{\bibfnamefont{P.}~\bibnamefont{Bruno}},
  \bibinfo{author}{\bibfnamefont{V.}~\bibnamefont{Dugaev}}, \bibnamefont{and}
  \bibinfo{author}{\bibfnamefont{M.}~\bibnamefont{Taillefumier}},
  \bibinfo{journal}{Phys.\ Rev.\ Lett.} \textbf{\bibinfo{volume}{93}},
  \bibinfo{pages}{096806} (\bibinfo{year}{2004}).

\bibitem[{\citenamefont{G\"obel et~al.}(2017)\citenamefont{G\"obel, Mook, Henk,
  and Mertig}}]{gobel2017THEskyrmion}
\bibinfo{author}{\bibfnamefont{B.}~\bibnamefont{G\"obel}},
  \bibinfo{author}{\bibfnamefont{A.}~\bibnamefont{Mook}},
  \bibinfo{author}{\bibfnamefont{J.}~\bibnamefont{Henk}}, \bibnamefont{and}
  \bibinfo{author}{\bibfnamefont{I.}~\bibnamefont{Mertig}},
  \bibinfo{journal}{Phys.\ Rev.\ B} \textbf{\bibinfo{volume}{95}},
  \bibinfo{pages}{094413} (\bibinfo{year}{2017}).

\bibitem[{\citenamefont{G{\"o}bel et~al.}(2017)\citenamefont{G{\"o}bel, Mook,
  Henk, and Mertig}}]{gobel2017QHE}
\bibinfo{author}{\bibfnamefont{B.}~\bibnamefont{G{\"o}bel}},
  \bibinfo{author}{\bibfnamefont{A.}~\bibnamefont{Mook}},
  \bibinfo{author}{\bibfnamefont{J.}~\bibnamefont{Henk}}, \bibnamefont{and}
  \bibinfo{author}{\bibfnamefont{I.}~\bibnamefont{Mertig}},
  \bibinfo{journal}{New J. Phys.} \textbf{\bibinfo{volume}{19}},
  \bibinfo{pages}{063042} (\bibinfo{year}{2017}).

\bibitem[{\citenamefont{Ndiaye et~al.}(2017)\citenamefont{Ndiaye, Akosa, and
  Manchon}}]{ndiaye2017topological}
\bibinfo{author}{\bibfnamefont{P.~B.} \bibnamefont{Ndiaye}},
  \bibinfo{author}{\bibfnamefont{C.~A.} \bibnamefont{Akosa}}, \bibnamefont{and}
  \bibinfo{author}{\bibfnamefont{A.}~\bibnamefont{Manchon}},
  \bibinfo{journal}{Phys.\ Rev.\ B} \textbf{\bibinfo{volume}{95}},
  \bibinfo{pages}{064426} (\bibinfo{year}{2017}).

\bibitem[{\citenamefont{Zang et~al.}(2011)\citenamefont{Zang, Mostovoy, Han,
  and Nagaosa}}]{zang2011dynamics}
\bibinfo{author}{\bibfnamefont{J.}~\bibnamefont{Zang}},
  \bibinfo{author}{\bibfnamefont{M.}~\bibnamefont{Mostovoy}},
  \bibinfo{author}{\bibfnamefont{J.~H.} \bibnamefont{Han}}, \bibnamefont{and}
  \bibinfo{author}{\bibfnamefont{N.}~\bibnamefont{Nagaosa}},
  \bibinfo{journal}{Phys.\ Rev.\ Lett.} \textbf{\bibinfo{volume}{107}},
  \bibinfo{pages}{136804} (\bibinfo{year}{2011}).

\bibitem[{\citenamefont{Jiang et~al.}(2017)\citenamefont{Jiang, Zhang, Yu,
  Zhang, Wang, Jungfleisch, Pearson, Cheng, Heinonen, Wang
  et~al.}}]{jiang2017direct}
\bibinfo{author}{\bibfnamefont{W.}~\bibnamefont{Jiang}},
  \bibinfo{author}{\bibfnamefont{X.}~\bibnamefont{Zhang}},
  \bibinfo{author}{\bibfnamefont{G.}~\bibnamefont{Yu}},
  \bibinfo{author}{\bibfnamefont{W.}~\bibnamefont{Zhang}},
  \bibinfo{author}{\bibfnamefont{X.}~\bibnamefont{Wang}},
  \bibinfo{author}{\bibfnamefont{M.~B.} \bibnamefont{Jungfleisch}},
  \bibinfo{author}{\bibfnamefont{J.~E.} \bibnamefont{Pearson}},
  \bibinfo{author}{\bibfnamefont{X.}~\bibnamefont{Cheng}},
  \bibinfo{author}{\bibfnamefont{O.}~\bibnamefont{Heinonen}},
  \bibinfo{author}{\bibfnamefont{K.~L.} \bibnamefont{Wang}},
  \bibnamefont{et~al.}, \bibinfo{journal}{Nature Physics}
  \textbf{\bibinfo{volume}{13}}, \bibinfo{pages}{162} (\bibinfo{year}{2017}).

\bibitem[{\citenamefont{Litzius et~al.}(2017)\citenamefont{Litzius, Lemesh,
  Kr{\"u}ger, Bassirian, Caretta, Richter, B{\"u}ttner, Sato, Tretiakov,
  F{\"o}rster et~al.}}]{litzius2017skyrmion}
\bibinfo{author}{\bibfnamefont{K.}~\bibnamefont{Litzius}},
  \bibinfo{author}{\bibfnamefont{I.}~\bibnamefont{Lemesh}},
  \bibinfo{author}{\bibfnamefont{B.}~\bibnamefont{Kr{\"u}ger}},
  \bibinfo{author}{\bibfnamefont{P.}~\bibnamefont{Bassirian}},
  \bibinfo{author}{\bibfnamefont{L.}~\bibnamefont{Caretta}},
  \bibinfo{author}{\bibfnamefont{K.}~\bibnamefont{Richter}},
  \bibinfo{author}{\bibfnamefont{F.}~\bibnamefont{B{\"u}ttner}},
  \bibinfo{author}{\bibfnamefont{K.}~\bibnamefont{Sato}},
  \bibinfo{author}{\bibfnamefont{O.~A.} \bibnamefont{Tretiakov}},
  \bibinfo{author}{\bibfnamefont{J.}~\bibnamefont{F{\"o}rster}},
  \bibnamefont{et~al.}, \bibinfo{journal}{Nature Physics}
  \textbf{\bibinfo{volume}{13}}, \bibinfo{pages}{170} (\bibinfo{year}{2017}).

\bibitem[{\citenamefont{Barker and Tretiakov}(2016)}]{barker2016static}
\bibinfo{author}{\bibfnamefont{J.}~\bibnamefont{Barker}} \bibnamefont{and}
  \bibinfo{author}{\bibfnamefont{O.~A.} \bibnamefont{Tretiakov}},
  \bibinfo{journal}{Phys.\ Rev.\ Lett.} \textbf{\bibinfo{volume}{116}},
  \bibinfo{pages}{147203} (\bibinfo{year}{2016}).

\bibitem[{\citenamefont{Zhang et~al.}(2016{\natexlab{a}})\citenamefont{Zhang,
  Zhou, and Ezawa}}]{zhang2016antiferromagnetic}
\bibinfo{author}{\bibfnamefont{X.}~\bibnamefont{Zhang}},
  \bibinfo{author}{\bibfnamefont{Y.}~\bibnamefont{Zhou}}, \bibnamefont{and}
  \bibinfo{author}{\bibfnamefont{M.}~\bibnamefont{Ezawa}},
  \bibinfo{journal}{Scientific Reports} \textbf{\bibinfo{volume}{6}},
  \bibinfo{pages}{24795} (\bibinfo{year}{2016}{\natexlab{a}}).

\bibitem[{\citenamefont{Fujita and Sato}(2017)}]{PhysRevB.95.054421}
\bibinfo{author}{\bibfnamefont{H.}~\bibnamefont{Fujita}} \bibnamefont{and}
  \bibinfo{author}{\bibfnamefont{M.}~\bibnamefont{Sato}},
  \bibinfo{journal}{Phys.\ Rev.\ B} \textbf{\bibinfo{volume}{95}},
  \bibinfo{pages}{054421} (\bibinfo{year}{2017}).

\bibitem[{\citenamefont{Jin et~al.}(2016)\citenamefont{Jin, Song, Wang, and
  Liu}}]{jin2016dynamics}
\bibinfo{author}{\bibfnamefont{C.}~\bibnamefont{Jin}},
  \bibinfo{author}{\bibfnamefont{C.}~\bibnamefont{Song}},
  \bibinfo{author}{\bibfnamefont{J.}~\bibnamefont{Wang}}, \bibnamefont{and}
  \bibinfo{author}{\bibfnamefont{Q.}~\bibnamefont{Liu}},
  \bibinfo{journal}{Applied Physics Letters} \textbf{\bibinfo{volume}{109}},
  \bibinfo{pages}{182404} (\bibinfo{year}{2016}).

\bibitem[{\citenamefont{Zhang et~al.}(2016{\natexlab{b}})\citenamefont{Zhang,
  Zhou, and Ezawa}}]{zhang2016magnetic}
\bibinfo{author}{\bibfnamefont{X.}~\bibnamefont{Zhang}},
  \bibinfo{author}{\bibfnamefont{Y.}~\bibnamefont{Zhou}}, \bibnamefont{and}
  \bibinfo{author}{\bibfnamefont{M.}~\bibnamefont{Ezawa}},
  \bibinfo{journal}{Nature Communications} \textbf{\bibinfo{volume}{7}},
  \bibinfo{pages}{10293} (\bibinfo{year}{2016}{\natexlab{b}}).

\bibitem[{\citenamefont{Buhl et~al.}(2017)\citenamefont{Buhl, Freimuth,
  Bl{\"u}gel, and Mokrousov}}]{buhl2017topological}
\bibinfo{author}{\bibfnamefont{P.~M.} \bibnamefont{Buhl}},
  \bibinfo{author}{\bibfnamefont{F.}~\bibnamefont{Freimuth}},
  \bibinfo{author}{\bibfnamefont{S.}~\bibnamefont{Bl{\"u}gel}},
  \bibnamefont{and}
  \bibinfo{author}{\bibfnamefont{Y.}~\bibnamefont{Mokrousov}},
  \bibinfo{journal}{physica status solidi (RRL)-Rapid Research Letters}
  \textbf{\bibinfo{volume}{11}}, \bibinfo{pages}{1700007}
  (\bibinfo{year}{2017}).

\bibitem[{\citenamefont{Do~Yi et~al.}(2009)\citenamefont{Do~Yi, Onoda, Nagaosa,
  and Han}}]{do2009skyrmions}
\bibinfo{author}{\bibfnamefont{S.}~\bibnamefont{Do~Yi}},
  \bibinfo{author}{\bibfnamefont{S.}~\bibnamefont{Onoda}},
  \bibinfo{author}{\bibfnamefont{N.}~\bibnamefont{Nagaosa}}, \bibnamefont{and}
  \bibinfo{author}{\bibfnamefont{J.~H.} \bibnamefont{Han}},
  \bibinfo{journal}{Phys.\ Rev.\ B} \textbf{\bibinfo{volume}{80}},
  \bibinfo{pages}{054416} (\bibinfo{year}{2009}).

\bibitem[{\citenamefont{Leonov and Mostovoy}(2015)}]{leonov2015multiply}
\bibinfo{author}{\bibfnamefont{A.}~\bibnamefont{Leonov}} \bibnamefont{and}
  \bibinfo{author}{\bibfnamefont{M.}~\bibnamefont{Mostovoy}},
  \bibinfo{journal}{Nature Comms.} \textbf{\bibinfo{volume}{6}},
  \bibinfo{pages}{8275} (\bibinfo{year}{2015}).

\bibitem[{\citenamefont{Hamamoto et~al.}(2015)\citenamefont{Hamamoto, Ezawa,
  and Nagaosa}}]{hamamoto2015quantized}
\bibinfo{author}{\bibfnamefont{K.}~\bibnamefont{Hamamoto}},
  \bibinfo{author}{\bibfnamefont{M.}~\bibnamefont{Ezawa}}, \bibnamefont{and}
  \bibinfo{author}{\bibfnamefont{N.}~\bibnamefont{Nagaosa}},
  \bibinfo{journal}{Phys.\ Rev.\ B} \textbf{\bibinfo{volume}{92}},
  \bibinfo{pages}{115417} (\bibinfo{year}{2015}).

\bibitem[{\citenamefont{Nagaosa et~al.}(2010)\citenamefont{Nagaosa, Sinova,
  Onoda, MacDonald, and Ong}}]{nagaosa2010anomalous}
\bibinfo{author}{\bibfnamefont{N.}~\bibnamefont{Nagaosa}},
  \bibinfo{author}{\bibfnamefont{J.}~\bibnamefont{Sinova}},
  \bibinfo{author}{\bibfnamefont{S.}~\bibnamefont{Onoda}},
  \bibinfo{author}{\bibfnamefont{A.}~\bibnamefont{MacDonald}},
  \bibnamefont{and} \bibinfo{author}{\bibfnamefont{N.}~\bibnamefont{Ong}},
  \bibinfo{journal}{Rev.\ Mod.\ Phys.} \textbf{\bibinfo{volume}{82}},
  \bibinfo{pages}{1539} (\bibinfo{year}{2010}).

\bibitem[{\citenamefont{Gradhand et~al.}(2012)\citenamefont{Gradhand, Fedorov,
  Pientka, Zahn, Mertig, and Gy{\"o}rffy}}]{gradhand2012first}
\bibinfo{author}{\bibfnamefont{M.}~\bibnamefont{Gradhand}},
  \bibinfo{author}{\bibfnamefont{D.}~\bibnamefont{Fedorov}},
  \bibinfo{author}{\bibfnamefont{F.}~\bibnamefont{Pientka}},
  \bibinfo{author}{\bibfnamefont{P.}~\bibnamefont{Zahn}},
  \bibinfo{author}{\bibfnamefont{I.}~\bibnamefont{Mertig}}, \bibnamefont{and}
  \bibinfo{author}{\bibfnamefont{B.}~\bibnamefont{Gy{\"o}rffy}},
  \bibinfo{journal}{Journal of Physics: Condensed Matter}
  \textbf{\bibinfo{volume}{24}}, \bibinfo{pages}{213202}
  (\bibinfo{year}{2012}).

\bibitem[{\citenamefont{Hatsugai}(1993{\natexlab{a}})}]{Hatsugai1993}
\bibinfo{author}{\bibfnamefont{Y.}~\bibnamefont{Hatsugai}},
  \bibinfo{journal}{Phys.\ Rev.\ B} \textbf{\bibinfo{volume}{48}},
  \bibinfo{pages}{11851} (\bibinfo{year}{1993}{\natexlab{a}}).

\bibitem[{\citenamefont{Hatsugai}(1993{\natexlab{b}})}]{Hatsugai1993a}
\bibinfo{author}{\bibfnamefont{Y.}~\bibnamefont{Hatsugai}},
  \bibinfo{journal}{Phys.\ Rev.\ Lett.} \textbf{\bibinfo{volume}{71}},
  \bibinfo{pages}{3697–3700} (\bibinfo{year}{1993}{\natexlab{b}}).

\bibitem[{\citenamefont{Yin et~al.}(2015)\citenamefont{Yin, Liu, Barlas, Zang,
  and Lake}}]{yin2015topological}
\bibinfo{author}{\bibfnamefont{G.}~\bibnamefont{Yin}},
  \bibinfo{author}{\bibfnamefont{Y.}~\bibnamefont{Liu}},
  \bibinfo{author}{\bibfnamefont{Y.}~\bibnamefont{Barlas}},
  \bibinfo{author}{\bibfnamefont{J.}~\bibnamefont{Zang}}, \bibnamefont{and}
  \bibinfo{author}{\bibfnamefont{R.~K.} \bibnamefont{Lake}},
  \bibinfo{journal}{Phys.\ Rev.\ B} \textbf{\bibinfo{volume}{92}},
  \bibinfo{pages}{024411} (\bibinfo{year}{2015}).

\bibitem[{\citenamefont{Henk and Schattke}(1993)}]{henk1993subroutine}
\bibinfo{author}{\bibfnamefont{J.}~\bibnamefont{Henk}} \bibnamefont{and}
  \bibinfo{author}{\bibfnamefont{W.}~\bibnamefont{Schattke}},
  \bibinfo{journal}{Computer Physics Communications}
  \textbf{\bibinfo{volume}{77}}, \bibinfo{pages}{69} (\bibinfo{year}{1993}).

\bibitem[{\citenamefont{B{\"o}dicker et~al.}(1994)\citenamefont{B{\"o}dicker,
  Schattke, Henk, and Feder}}]{bodicker1994interface}
\bibinfo{author}{\bibfnamefont{A.}~\bibnamefont{B{\"o}dicker}},
  \bibinfo{author}{\bibfnamefont{W.}~\bibnamefont{Schattke}},
  \bibinfo{author}{\bibfnamefont{J.}~\bibnamefont{Henk}}, \bibnamefont{and}
  \bibinfo{author}{\bibfnamefont{R.}~\bibnamefont{Feder}},
  \bibinfo{journal}{Journal of Physics. Condensed Matter}
  \textbf{\bibinfo{volume}{6}}, \bibinfo{pages}{1927} (\bibinfo{year}{1994}).

\bibitem[{\citenamefont{Woo et~al.}(2017)\citenamefont{Woo, Song, Zhang, Zhou,
  Ezawa, Finizio, Raabe, Choi, Min, Koo et~al.}}]{woo2017current}
\bibinfo{author}{\bibfnamefont{S.}~\bibnamefont{Woo}},
  \bibinfo{author}{\bibfnamefont{K.~M.} \bibnamefont{Song}},
  \bibinfo{author}{\bibfnamefont{X.}~\bibnamefont{Zhang}},
  \bibinfo{author}{\bibfnamefont{Y.}~\bibnamefont{Zhou}},
  \bibinfo{author}{\bibfnamefont{M.}~\bibnamefont{Ezawa}},
  \bibinfo{author}{\bibfnamefont{S.}~\bibnamefont{Finizio}},
  \bibinfo{author}{\bibfnamefont{J.}~\bibnamefont{Raabe}},
  \bibinfo{author}{\bibfnamefont{J.~W.} \bibnamefont{Choi}},
  \bibinfo{author}{\bibfnamefont{B.-C.} \bibnamefont{Min}},
  \bibinfo{author}{\bibfnamefont{H.~C.} \bibnamefont{Koo}},
  \bibnamefont{et~al.}, \bibinfo{journal}{arXiv preprint arXiv:1703.10310}
  (\bibinfo{year}{2017}).

\bibitem[{\citenamefont{Mook et~al.}(2017)\citenamefont{Mook, G{\"o}bel, Henk,
  and Mertig}}]{mook2017magnon}
\bibinfo{author}{\bibfnamefont{A.}~\bibnamefont{Mook}},
  \bibinfo{author}{\bibfnamefont{B.}~\bibnamefont{G{\"o}bel}},
  \bibinfo{author}{\bibfnamefont{J.}~\bibnamefont{Henk}}, \bibnamefont{and}
  \bibinfo{author}{\bibfnamefont{I.}~\bibnamefont{Mertig}},
  \bibinfo{journal}{Phys.\ Rev.\ B} \textbf{\bibinfo{volume}{95}},
  \bibinfo{pages}{020401} (\bibinfo{year}{2017}).

\end{thebibliography}
\bibliographystyle{apsrev}


\end{document}